\documentclass[prd,twocolumn,nofootinbib,preprintnumbers,floatfix]{revtex4}
\usepackage[utf8]{inputenc}
\usepackage{amsfonts,amsmath,amssymb}
\usepackage{graphicx}
\usepackage{color}
\usepackage[plainpages=false, colorlinks=true, anchorcolor=blue, linkcolor=blue, citecolor=blue, bookmarks=false]{hyperref}
\usepackage{natbib}
\usepackage{enumitem}
\usepackage{subcaption}
\newcommand{\rthis}[1]{\textcolor{black}{#1}}
\begin{document}
\newcommand{\apjl}{Astrophys. J. Lett.}
\newcommand{\apjs}{Astrophys. J. Suppl. Ser.}
\newcommand{\aap}{Astron. \& Astrophys.}
\newcommand{\aj}{Astron. J.}
\newcommand{\araa}{Ann. Rev. Astron. Astrophys. } 
\newcommand{\aapr}{Astronomy and Astrophysics Review}
\newcommand{\mnras}{Mon. Not. R. Astron. Soc.}
\newcommand{\apss} {Astrophys. and Space Science}
\newcommand{\jcap}{JCAP}
\newcommand{\pasj}{PASJ}
\newcommand{\pasa}{Pub. Astro. Soc. Aust.}
\newcommand{\physrep}{Physics Reports}
\newcommand{\ssr}{Space Science Reviews}

\title{Scaling relations for dark matter core density and radius from Chandra X-ray cluster sample}
\author{Gopika \surname{K.}}\altaffiliation{E-mail:ph19resch01001@iith.ac.in}

\author{Shantanu  \surname{Desai}}  
\altaffiliation{E-mail: shntn05@gmail.com}

\begin{abstract}
A large number of studies have found that the dark matter surface density, given by the product of the dark matter core radius ($r_c$) and core density ($\rho_c$) is approximately constant for a wide range of galaxy systems. However, there has been only one systematic study  of this {\it ansatz} for galaxy clusters by Chan~\cite{Chan}, who  found that the surface density for  clusters is not constant and $\rho_c \sim r_c^{-1.46}$. We carry out this test for an   X-ray sample of 12 relaxed clusters from Chandra observations, studied by Vikhlinin et al~\cite{2006ApJ...640..691V}, implementing the same procedure as  in Ref.~\cite{Chan}, but also accounting for the gas and star mass. We find that $\rho_c \propto r_c^{-1.08 \pm 0.055}$, with  an intrinsic scatter of about 18\%. Therefore,   the dark matter surface density for our cluster data  shows  deviations from a constant value  at only about 1.4$\sigma$. 

\end{abstract}

\affiliation{Department of Physics, Indian Institute of Technology, Hyderabad, Telangana-502285, India}
\maketitle

\section{Introduction}
The current concordance ($\Lambda$CDM) cosmological model  consisting of 25\% cold dark matter and 70\% dark energy, agrees very well with Planck CMB and large scale structure observations~\cite{Planck2018}. However, at  scales smaller than about 1~Mpc, the cold dark matter paradigm runs into a number of problems such as the core/cusp problem, missing satellite problem (although see ~\cite{Peter}), too big to fail problem, satellites plane problem etc.~(See Refs.~\cite{Bullock,Weinberg,Tulin}
for recent reviews on this subject). We also note that some  of these problem can be ameliorated using various  baryonic physics effects~\cite{Oh,Martizzi,Mori,Keres}.
At a more fundamental level, another issue with the $\Lambda$CDM model is that there is no laboratory evidence for any cold dark matter  candidate~\cite{Merritt}. Furthermore, LHC or other particle physics experiments have yet to find experimental evidence for theories beyond the Standard Model of Particle Physics, which predict such cold dark matter candidates~\cite{Merritt}. Therefore, a large number of theoretical alternatives to  $\Lambda$CDM model have been proposed, and a variety of observational tests devised to test these myriad alternatives.

An intriguing observational result discovered more than a decade ago is that
the dark matter halo surface density is constant,  for a wide variety of systems spanning over 18 orders in blue magnitude  for a diverse suite of galaxies, such as  spiral galaxies, low surface brightness galaxies,  dwarf spheroidal satellites of Milky way~\cite{Kormendy,Donato,Gentile,Walker,Salucci,Hartwick,Kormendy14,Chiba,Burkert15,Salucci19} etc. See however Refs.~\cite{Boyarsky,Napolitano,Cardone10,DelPopolo12,Saburova,DelPopolo17,DelPopolo20} and references therein, which dispute these claims and argue  for a  mild dependence of the dark matter surface density with halo mass and other galaxy properties. These results for a constant dark matter surface density were obtained by fitting the dark matter distribution in these systems to  a cored profile, either  Burkert~\cite{Burkert95}, pseudo-isothermal profile~\cite{Kormendy}, or a simple isothermal sphere~\cite{Spano}. All these cored profiles can be parameterized by a  central density ($\rho_c$) and core radius ($r_c$); and the halo surface density is defined as the product of  $\rho_c$ and $r_c$. The existence of a constant dark matter surface density was found to be  independent of which cored profile was used~\cite{Donato}. Alternately, some groups  have also calculated a variant of  the above dark matter halo density, which has been referred to as  the dark matter column density~\cite{Boyarsky,DelPopolo12}~\footnote{See Eq. 1 of Ref.~\cite{Boyarsky} for the definition of dark matter column density.}, whose value remains roughly invariant with respect to the choice of the dark matter profile. This column density is equivalent to the product of  $\rho_c$ and $r_c$ for a Burkert profile~\cite{DelPopolo12}, and provides  a more precise value of the surface density for non-cored profiles, such as the widely used NFW profile~\cite{NFW}.
The best-fit values for the dark matter  surface density  for single galaxy systems  using the latest observational data is given by $\log  (\rho_c r_c)=(2.15 \pm 0.2) M_{\odot} pc^{-2}$~\cite{Salucci19}.

A large number of theoretical explanations have been proposed to explain the constancy of dark matter halo density. Within the standard $\Lambda$CDM model, some explanations include: transformation of cusps to cores due to dynamical feedback processes~\cite{Ogiya},   self-similar secondary infall model~\cite{Delpopolo09,Boyarsky,DelPopolo12,DelPopolo17}, dark matter-baryon interactions~\cite{Famaey}, non-violent relaxation of galactic haloes~\cite{Baushev}. Some explanations beyond $\Lambda$CDM include ultralight scalar dark matter~\cite{Matos}, super-fluid dark matter~\cite{Berezhiani}, self-interacting dark matter~\cite{Loeb,Kaplinghat,Bondarenko,Tulin},  MOND~\cite{Milgrom}, etc. A constant halo surface density is also in tension with fuzzy dark matter models~\cite{Burkert20}.

It behooves us to test the same relation for galaxy clusters. Galaxy clusters are the most massive collapsed objects in the universe and are a wonderful laboratory for a wide range of topics from cosmology  to galaxy evolution~\cite{Mantz,Vikhlininrev}. In the last two decades a large number of new galaxy clusters have been discovered through dedicated optical, X-ray, and SZ surveys, which have provided a wealth of information on Astrophysics and Cosmology.  However, tests of the constancy of dark matter surface density for galaxy clusters have been very few.

The first such study for galaxy clusters was done by Boyarsky et al~\cite{Boyarsky09}, who  used the dark matter profiles from literature   for 130 galaxy clusters and showed that the dark matter  column density ($S$) goes as  $S \propto M_{200}^{0.21}$, where $M_{200}$ is the mass within the density contrast at $\Delta=200$~\cite{White}, where the density contrast ($\Delta$) is defined with respect to the   critical density.
 Hartwick~\cite{Hartwick}  used  the generalized NFW  profile~\cite{NFW} fits in Ref.~\cite{Newman} (using strong and weak lensing data) for the Abell 611 cluster,   and found that $\rho_c r_c = 2350 M_{\odot} pc^{-2}$.  This is about twenty times larger than the corresponding value obtained for galaxies~\cite{Salucci19}. 
Lin and Loeb~\cite{Loeb} estimated  $\rho_c r_c \approx 1.1 \times 10^3 M_{\odot} pc^{-2}$ for the Phoenix cluster, using multi-wavelength data  obtained by the SPT collaboration~\cite{SPT2012}. Using a model for self-interacting dark matter including annihilations, they also predicted 
the following relation between the surface density and  $M_{200}$~\cite{Loeb} 
\begin{equation}
\rho_c r_c = 41 M_{\odot} pc^{-2} \times \left(\frac{M_{200}}{10^{10}M_{\odot}}\right)^{0.18}
\label{loebeq}
\end{equation}
Del Popolo et al also predicted~\cite{DelPopolo12,DelPopolo17} a similar relation between the dark matter column density and $M_{200}$, within the context of a secondary infall model~\cite{Delpopolo09} valid for cluster scale haloes
\begin{equation}
\log (S)= 0.16  \log \left(\frac{M_{200}}{10^{12}M_{\odot}}\right)+2.23
\label{delpopoloeq}
\end{equation}

The first systematic study of the correlation between $\rho_c$ and $r_c$ for an X-ray selected cluster sample, and without explictly assuming any dark matter density model,  was done by Chan~\cite{Chan} (C14, hereafter). C14 first considered the X-ray selected HIFLUGCS cluster catalog consisting of ASCA and ROSAT
observations~\cite{Chen07}. They considered  106 relaxed clusters from this catalog.
From the hydrostatic equilibrium equation and parametric models  for the gas density and temperature profiles, the total mass ($M(r)$) was obtained as a function of radius. The total density as a function of radius ($\rho(r)$) was then obtained from the total mass, assuming spherical symmetry.

One premise in C14 is that the total mass is dominated by the dark matter contribution, while the stellar  and gas mass  can be ignored. $\rho_c$ was obtained from extrapolating the dark matter density distribution to  $r=0$.  The core radius was obtained by finding the radius ($r$) at which $\rho (r)=\rho_c/4$. This emulates the definition of $r_c$ in the Burkert profile~\cite{Burkert95}.  Therefore, the estimate of core density and radius  was done in C14 without explicitly positing any dark matter profile.
We  note that from weak and strong lensing observations, galaxy clusters are estimated to have cored or shallower than cuspy NFW dark matter profiles~\cite{Newman12,DelPopolo14}. However, these results have been disputed~\cite{Schaller}, and some works have also found evidence for cuspy haloes in clusters~\cite{Caminha}. Therefore, there is no consensus on this issue~\cite{Andrade}. 
Nevertheless, no explicit assumptions about the  dark matter profile was made in C14, while obtaining the dark matter core density and radius, although the gas density models used for deducing this,  were initially motivated from assuming isothermality and  a  King profile for the total  mass distribution in galaxy clusters.
We also note that in some models, for example the cusp to core transformation  model~\cite{Ogiya} or the self-interacting dark matter with annihilations~\cite{Loeb}, the product of the core density and core radius for the cored profile is same as the product of scale density and scale radius of cuspy NFW-like profiles. 

In their analysis, C14 used two different density profiles (single-$\beta$ and double-$\beta$ model) for the gas density. They also did separate fits for  both the cool-core and the non cool-core clusters. Using the double-$\beta$ model, they obtained $\rho_c \propto  r_c^{-1.46 \pm 0.16}$ for the
HIFLUGCS sample. Results from fits with other profiles for the same sample as well as other samples can be found in C14.
Therefore, their result shows that the dark matter surface density is not constant for clusters. C14 also
carried out a similar analysis on the LOCUSS cluster sample analyzed in Shan et al~\cite{Shan} and found that $\rho_c \propto  r_c^{-1.64 \pm 0.10}$. Therefore, these results indicate that unlike single-galaxy systems, the dark matter surface density is not constant for galaxy clusters
and is about an  order of magnitude larger than for single galaxy systems.

We now implement the procedure recommended in C14 to determine $\rho_c$ and $r_c$ for  a catalog of 12 galaxy clusters, selected using pointed X-ray and archival ROSAT observations by Vikhlinin et al~\cite{2006ApJ...640..691V} (V06, hereafter). Detailed parametric profiles for gas density and temperature profiles have been compiled by V06. This cluster sample has been used to constrain a  plethora of modified gravity theories and also to test non-standard alternatives to $\Lambda$CDM model ~\cite{2014PhRvD..89j4011R,Khoury17,Bernal,Ng,Hodson17}. We have also  previously used this sample to constrain the graviton mass~\cite{Gupta1} as well as to assess the importance of relativistic corrections to hydrostatic mass estimates~\cite{Gupta2}. Our work improves upon C14 in that, we account for  the baryonic mass distribution while estimating the dark matter halo properties.

The outline of this manuscript is as follows. We describe the V06 cluster sample and associated models for the density and temperature profile in Sect.~\ref{sec:data}. Our analysis and results for the relation between core radius and density can be found in Sect.~\ref{sec:analysis}. Comparison with various theoretical scenarios is discussed in Sect.~\ref{sec:theory}. We also test for dependence vs $M_{200}$ in Sect.~\ref{sec:m200}. We conclude in Sect.~\ref{sec:conclusions}.

\section{Details of Chandra X-ray sample}
\label{sec:data}
V06  (See also Ref.~\cite{Vikhlinin05}) derived  density and temperature profiles for a  total of 13 nearby relaxed galaxy clusters (A133, A262, A383, A478, A907, A1413, A1795, A1991, A2029, A2390, MKW4, RXJ1159+55531, USGC 2152) using measurements  obtained from  pointed  observations with the  Chandra X-ray satellite.  The redshifts of these clusters range approximately upto $z=0.2$. These measurements extended up to very large radii  of about $r_{500}$ for some of the clusters. For lower redshift clusters, because of Chandra's limited field of view, archival ROSAT observations between 0.7-2.0 keV  were used as a second independent data set to model the gas density distribution at large radii. The list of clusters for which ROSAT archival observations were used in conjunction with pointed Chandra observations include A133, A262,
A478, A1795, A1991, A2029, and MKW 4.
The redshifts of these clusters range approximately upto $z=0.2$. These measurements extended up to very large radii  of about $r_{500}$ for some of the clusters.  The typical Chandra exposure times ranged from 30-130 ksecs. The temperatures span the range between  1 and 10 keV and masses from $(0.5-10) \times 10^{14} M_{\odot}$. V06 provided analytical estimates for the 3-D gas density and temperature profiles used to reconstruct the masses. The accuracy of mass reconstruction, tested with simulations was estimated to be within a few percent. From this sample of 13 clusters, we skipped  USGC 2152  for our analysis, as all the relevant data was not available to us. More details about this cluster sample can be found in V06.

We now describe the  three-dimensional models proposed by V06, for the gas density and temperature projected along the line of sight. These models can fit the observed X-ray surface brightness and projected temperature profiles. These parametric models are then used to derive the total gravitating mass of the clusters.
\subsection{Gas Density Model}
The analytic expression used for the three-dimensional gas
density distribution is a modified version of the  single-$\beta$-model~\citep{1978A&A....70..677C}.  These modifications were  introduced to account for some additional  features in the observed X-ray emission, such as a power-law cusp at the center and a steepening of the  X-ray brightness for $r>0.3 r_{200}$.
To model the core region, a second $\beta$-profile was added  to increase the modeling freedom.
This  modified emission $\beta$-profile is then defined as,
\begin{eqnarray}
  n_pn_e=n_0^2\frac{(r/r_c)^{-\alpha}}{(1+r^2/r_c^2)^{3\beta-\alpha/2}}\frac{1}{(1+r^\gamma/r_s^\gamma)^{\epsilon/\gamma}}\nonumber\\+\frac{n_{02}^2}{(1+r^2/r_{c2}^2)^{3\beta_2}}
      \label{eq:eq3}
\end{eqnarray}
where $n_p$ and $n_e$ denote the number density of protons and electrons respectively. The model in Eq.~\ref{eq:eq3} can independently fit the inner and outer cluster regions, and all the clusters in V06 sample were adequately fit with a fixed  value of $\gamma$ equal to  three.
A detailed description of all the other parameters in Eq.~\ref{eq:eq3},  as well as their values can be found in V06. The density and radius related constants used to parametrize Eq.~\ref{eq:eq3}, scale with the dimensionless Hubble parameter $h$ as 
$h^{1/2}$ and $h^{-1}$ respectively. V06 has used $H_0$=72 km/sec/Mpc.
We have used the same values for the best-fit parameters of all the terms in Eq.~\ref{eq:eq3}  as in  V06, where they can be found in Table 2. V06 however notes that the parameters in Eq.~\ref{eq:eq3} are correlated, leading to degeneracies. However, the analytical expression in Eq.~\ref{eq:eq3} can adequately model the X-ray brightness for all clusters. No errors for individual fit parameters have been provided in V06. The statistical uncertainties have been estimated using Monte Carlo simulations  and found to be less than 9\%~\cite{2006ApJ...640..691V}.

From the gas particle number density profile given by Eq.~\ref{eq:eq3}, the gas mass density can be obtained by assuming a cosmic plasma with primordial Helium abundance and abundances of
heavier elements $Z=0.2Z_\odot$ as,
\begin{equation}
\rho_g=1.624m_p(n_pn_e)^{1/2}
    \label{eq:eq4}
\end{equation}

\subsection{Temperature Profile Model}
To calculate the total dynamical mass, we need the three-dimensional temperature radial profile, whereas X-ray observations can only constrain the projected two-dimensional profile.
The reconstructed temperature profile in V06 consists of two different functions, one to model the central part and another to model the region outside the central cooling zone.
 A broken power law is used to model  the temperature  outside the central cooling region and is given as,
\begin{equation}
t(r)=\frac{(r/r_t)^{-a}}{\big[1+(r/r_t)^b\big]^{c/b}}
\label{eq:eq5}
\end{equation}
The temperature decline in the central region can be expressed as~\citep{2001MNRAS.328L..37A},
\begin{equation}
t_{cool}(r)=\frac{(x+T_{min}/T_0)}{x+1}, \quad x=\bigg(\frac{r}{r_{cool}}\bigg)^{a_{cool}}
\label{eq:eq6}
\end{equation}
The three-dimensional temperature profile of the cluster is given:
\begin{equation}
T_{3D}(r)=T_0t_{cool}(r)t(r)
\label{eq:eq7}
\end{equation}
Due to the large number (nine) of free parameters  in the model, Eq.~\ref{eq:eq7} can adequately describe any smooth trend in temperature profile. However, there are also degeneracies between the parameters.
Therefore, again no errors were provided in V06 for the individual parameters describing the temperature profile. For doing the fits, the temperature data below an inner cutoff radius (usually between 10-40 kpc, with exact values tabulated in V06)  was excluded  because the intracluster medium is expected to be multiphase at such small radii~\cite{2006ApJ...640..691V}. 
The best-fit parameters for this model  can be found  in Table 3 of V06. As can be seen in Figs 3-14 of V06, the analytical temperature agrees very well with the projected temperatures for all clusters for all radii greater than the inner cutoff radius. 
\subsection{Mass and Density Profile in Clusters}
\label{sec:procedure}
The total mass of the galaxy cluster can be derived through hydrostatic equilibrium equation, given the temperature and gas density models~\cite{Mantz},
\begin{equation}
M(r)=-\frac{kT(r)r}{G\mu m_p}\bigg(\frac{d \ln \rho_g}{d \ln r} + \frac{d \ln T}{d \ln r}\bigg)
\label{eq:eq8}
\end{equation}
where  $M(r)$ is the mass within radius $r$; $T$ and $\rho_g$ denote the gas temperature and density; $\mu$ is the mean molecular weight equal to 0.5954 as in V06, and $m_p$ is the mass of the proton.  

\par We can estimate the  total dark matter mass distribution  by subtracting the gas  and stellar mass from the total mass, given by  Eq. ~\ref{eq:eq8}. The gas mass can be simply obtained by assuming spherical symmetry and integrating the gas density profile ($\rho_g(r)$ from Eq.~\ref{eq:eq4})
\begin{equation}
    M_{gas}=\int 4\pi r^2 \rho_g(r)dr
    \label{eq:eq9}
\end{equation}
The gas mass  and total mass  can hence be robustly determined for each cluster using Eq.~\ref{eq:eq9} and ~\ref{eq:eq8}. V06 however cautions that A2390 reveals X-ray cavities at small radii due to AGN activity~\cite{Allen01}. Therefore, its gas mass will be overestimated and the total mass underestimated. We do not account for this in our estimates of the dark matter core radius and density for this cluster.

To calculate the stellar mass at any radius ($M_{star} (r)$), we first estimate the star mass  at $r=r_{500}$, where $r_{500}$ is the radius at which the overdensity is equal to 500, where
overdensity is again defined  relative to the critical density for an Einstein-DeSitter universe at the cluster redshift $z$.
This star mass at $r_{500}$ was estimated using the empirical relation proposed in Ref.~\cite{ytlin}, assuming $H_0$=71 km/sec/Mpc
\begin{equation}
\frac{M_{star}(r=r_{500})}{10^{12}M_\odot}\approx (1.8 \pm 0.1) \bigg(\frac{M_{500}}{10^{14}M_\odot}\bigg)^{0.71 \pm 0.04 }
   \label{eq:eq10}
\end{equation}
The above empirical relation was  determining by comparing $M_{500}$ and star mass for a sample of about 100 clusters in the redshift range $z=0.1-0.6$. $M_{500}$ was determined using the $Y_X$-$M_{500}$ scaling relation from ~\cite{Kravtsov}. The stellar mass was estimated from the  WISE or 2MASS (depending on the redshift range) luminosity and the Bruzual-Charlot stellar population synthesis models~\cite{Charlot}. Lin et al also find no redshift evolution of this empirical relation~\cite{ytlin}.
To estimate the star mass at $r_{500}$, for each cluster, we used $r_{500}$ tabulated for each cluster in V06. $r_{500}$  was estimated for every cluster using the cosmological parameters used in V06 : $\Omega_M=0.3,\Omega_{\Lambda}=0.7$ and $h=0.72$. 
From Eq.~\ref{eq:eq10}, one can estimate the star mass at any radius, by assuming an isothermal profile~\cite{2014PhRvD..89j4011R}
\begin{equation}
M_{star} (r)=\left(\frac{r}{r_{500}}\right)M_{star}(r=r_{500}) 
\end{equation}
Alternately, the stellar mass can also be estimated using the stellar-to-gas mass relation obtained in Chiu et al~\cite{Chiu18}, as used in Ref.~\cite{Tian}.
However, since the stellar mass contribution to the total mass is negligible, this will not make a large difference to the final result.

Therefore, once we estimate the star and gas mass, we can determine the  total dark matter mass distribution ($M_{DM} (r)$) at any radius ($r$) by subtracting  the gas and star mass from the total mass distribution ($M(r)$) calculated in Eq.~\ref{eq:eq8}:
\begin{equation}
M_{DM} (r)=M (r)-M_{gas} (r)-M_{star} (r)
   \label{eq:eq11} 
\end{equation}

From Eq.~\ref{eq:eq11}, the density profile of the dark matter halo can be easily calculated by assuming spherical symmetry:
 \begin{equation}
\rho_{DM} (r) =\frac{1}{4\pi r^2}\frac{dM_{DM}}{dr}
\label{eq:eq12} 
\end{equation}
\par 
To obtain $\rho_c$ and $r_c$ from $\rho_{DM} (r)$, we follow the same prescription as in C14,  which we now describe.
To recap, $\rho_c$ is estimated from the dark matter density at the centre of the cluster. Therefore, similar to C14, we extrapolated our dark matter density profile $\rho_{DM} (r)$ (obtained from Eq.~\ref{eq:eq12}) to $r=0$ in order to  calculate $\rho_c$. The core radius $r_c$  was  estimated by determining the radius at which the local dark matter density (defined in Eq.~\ref{eq:eq12}) reaches a quarter of its central value. As mentioned earlier, this  is how $r_c$ is defined in the Burkert profile~\citep{Burkert95,burkert2000structure,Gentile}. 

In this work and similar to C14,  we also estimate $\rho_c$ and $r_c$ in a model-independent way without explicitly positing any dark matter profile. However, we should point out that the single-$\beta$ profile (used  in C14 and also the starting point for Eq.~\ref{eq:eq3}) for the gas distribution, was originally derived from the equation of hydrostatic equilibrium
for an isothermal gas, and assuming that the total cluster matter distribution is based on King profile~\cite{1978A&A....70..677C}. Both C14 and this work have used an augmented version of this (double and modified  $\beta$-profile)  to fit the X-ray observations.  Nevertheless, the same 
$\beta$-profile is also usually used in testing modified gravity theories or alternatives to $\Lambda$CDM with galaxy cluster observations~\cite{2014PhRvD..89j4011R,Gupta2,Hodson17,Khoury17}. However, a truly {\it ab-initio}
determination of the dark matter halo surface density is not possible.


We note that we have assumed spherical symmetry in the calculation of gas and total mass.
Although we expect   galaxy clusters  to be  triaxial~\cite{Jing,Limousin,Battaglia}, one usually resorts to spherically averaged measurements of clusters, including  in V06 and C14, and also for all other tests of modified gravity using relaxed clusters. This is due to the simplicity and the assumption that errors due to  non-spherical effects are small. Furthermore, the intrinsic shapes and orientation of clusters are not directly observable and  reconstructing them  from X-ray images, which are inherently 2-D in nature can be very challenging~\cite{Limousin,Khatri}. Nevertheless, a large number of groups have investigated the errors in the determination of the total mass due to spherical symmetry assumption~\cite{Piffaretti,Gavazzi,Clowe,Battaglia,Buote1,Buote2,Limousin} (and references therein). The actual error depends on the intrinsic shape of the cluster and orientation along the line of sight. All these groups find that spherical averaging causes errors $\leq 5\%$  for all different intrinsic shapes and viewing orientations~\cite{Piffaretti,Gavazzi,Clowe,Battaglia,Buote1,Buote2,Limousin}. We therefore expect the same for the clusters in V06 sample.

We  now use Eq.~\ref{eq:eq12} to determine 
$\rho_c$ and $r_c$ for  the 12 clusters in V06.
Errors in the observed gas temperature   at a fixed number of radii have also been provided  in V06 and made available to us (A. Vikhlinin, private communication). These were used to propagate the errors in the values of $\rho_c$ and $r_c$, by varying the temperatures within the $1\sigma$ error bars. As no errors in the  parameters of the gas and temperature density profile  have been provided
in V06, these were not used to calculate the 
errors in $\rho_c$ and $r_c$.

\section{Analysis and Results}
\label{sec:analysis}
\subsection{Determination of scaling relations}
 The resulting values of  $\rho_c$ and $r_c$  along with $1\sigma$ error bars for each of the 12 clusters estimated using the procedure outlined in Sect.~\ref{sec:procedure}  can be found in Table \ref{tab:table4}. We  note that our values for $\rho_c$ and $r_c$ are of the same order of magnitude as for other galaxy cluster systems estimated in C14. 
 Our estimated dark matter surface density is about an order of magnitude larger than that found for galaxy systems~\cite{Donato,Salucci19}. 
 
\begin{figure*}
    \includegraphics{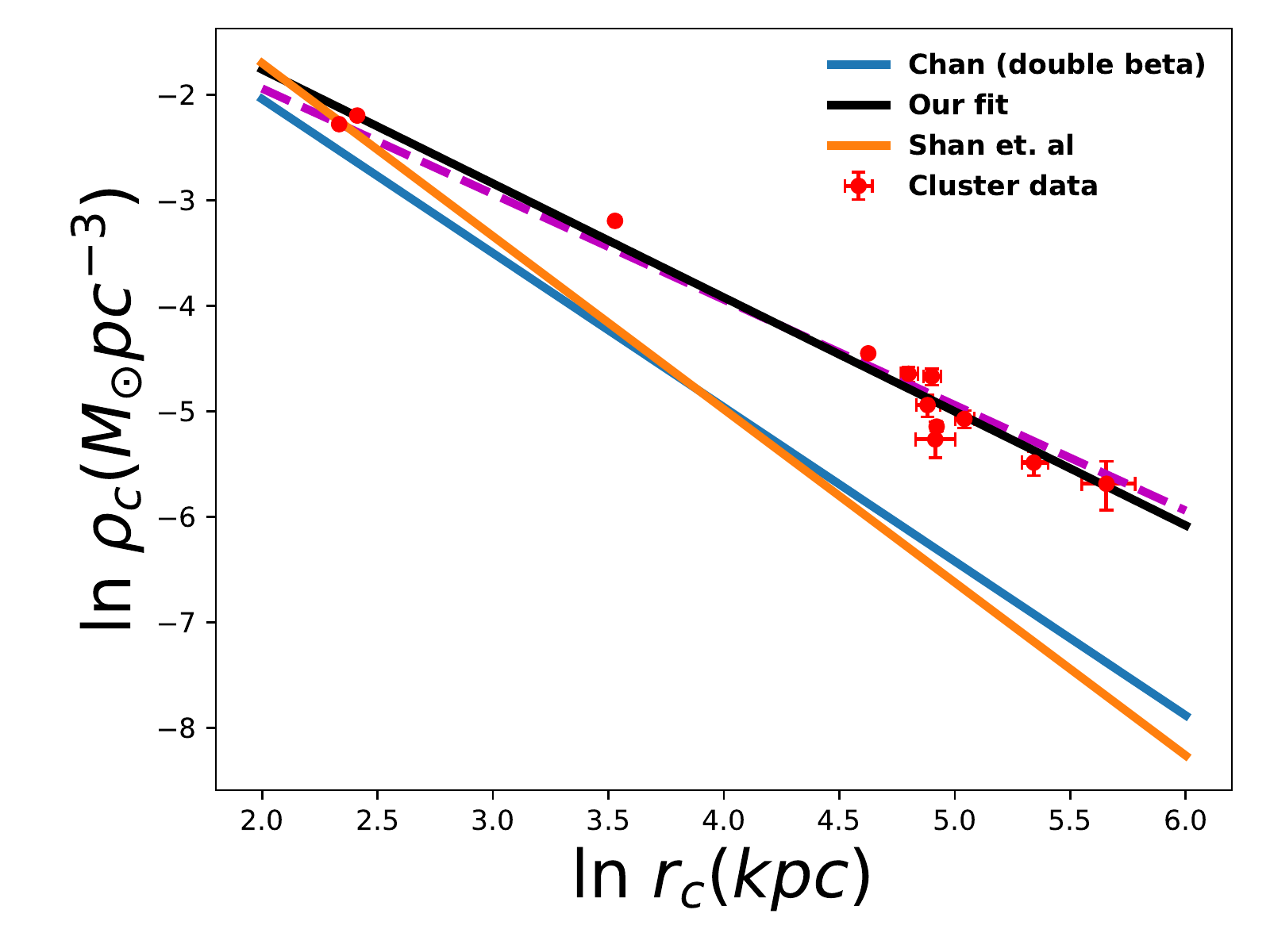}
    \caption{$\ln \rho_c$ versus $\ln r_c$ from V06 cluster sample~\cite{2006ApJ...640..691V}. The units for $\rho_c$ and $r_c$ are in $M_\odot pc^{-3}$ and kpc respectively. The black line represents the fitted line of our analysis ($\rho_c \propto r_c^{-1.08}$, and can be compared with the dashed magenta line, which has a slope equal to -1. The blue and 
 orange    lines show the slope  determined by Chan~\cite{Chan} for ROSAT catalog and the Shan et al~\cite{Shan} sample respectively.   A summary of all results in literature can be found in Table~\ref{tab:table4}.}
    \label{fig:f3}
    \end{figure*}

Figure \ref{fig:f3} shows the log $\rho_c$ versus log $r_c$ plot, and we  observe a tight scaling relation between the two. We also find that $\rho_c$ is inversely proportion to $r_c$ in agreement with C14. To determine the scaling relation between the two, we perform a linear regression ($y=mx+c$)  in log-log space. Here, $y=\ln \rho_c$ and $x=\ln r_c$.  Unlike C14, we also allow for an intrinsic scatter ($\sigma_{int}$) in the linear fit. This intrinsic scatter is treated as a free parameter and is added in quadrature to the observational uncertainties ($\sigma_y$ and $\sigma_x$). It can be determined along with the slope and intercept by maximizing the log-likelihood~\cite{Tian,Hoekstra}.
The log-likelihood function ($\ln L$) can be written as,
\begin{eqnarray}
-2\ln L &=& \large{\sum_i} \ln 2\pi\sigma_i^2 + \large{\sum_i} \frac{[y_i-(mx_i+c)]^2}{\sigma_i^2}
\label{eq:eq13}  \\
\sigma_i^2 &=& \sigma_{y_i}^2+m^2\sigma_{x_i}^2+\sigma_{int}^2
\end{eqnarray}

The  maximization of the log-likelihood was done using the  {\tt emcee} MCMC sampler~\cite{emcee} with uniform priors. Our best-fit value for the scaling relation is as follows:

\begin{eqnarray}
\ln\bigg(\frac{\rho_c}{M_{\odot} pc^{-3}}\bigg)=(-1.08^{+0.06}_{-0.05}) \ln\bigg(\frac{r_c}{kpc}\bigg)\nonumber\\
+(0.4^{+0.24}_{-0.25})  
\label{eq:eq14} 
\end{eqnarray}
with an intrinsic scatter, $\sigma_{int}=18.5^{+4.4}_{-6.5} \%$.
Therefore, we obtain  a much shallower slope for the core density-radius scaling relation compared to the slope obtained in C14, who found $\rho_c \sim r_c^{-1.46}$ for the HIFLUGCS sample, after assuming a double-beta profiles. Our results show deviations from a  constant dark matter surface density at only about 1.3-1.4$\sigma$. A comparison of our result with the previous fits carried out  in C14 can be found in Table~\ref{tab:summary}. As we can  see all the fits done in C14 show a much steeper slope than our result.


\begin{table}[h]
\begin{ruledtabular}
\begin{tabular}{ccc}
Cluster & $\rho_c$ & $r_c$ \\
& $10^{-3}M_{\odot} pc^{-3}$ & kpc \\\hline
 A133 &   11.68$\substack{ +0.02 \\ -0.02 }$ & 102.01$\substack{+0.08 \\ -0.11 }$ \\
 A262 &  5.17$\substack{ +0.87\\ -0.89}$ & 136.36$\substack{+5.40 \\-5.49 }$\\
A383 & 9.63$\substack{+0.62 \\ -0.78}$ & 121.45$\substack{ +3.95\\-4.94 }$\\
A478  & 3.39$\substack{+0.72 \\ -0.84}$ & 286.14$\substack{+30.41 \\-35.62 }$ \\
A907 & 4.15$\substack{ +0.42\\ -0.51 }$ & 208.96$\substack{+10.66 \\-12.98 }$\\
A1413 &  6.27$\substack{ +0.49\\ -0.53 }$ & 154.68$\substack{+6.06 \\ -6.61 }$\\
A1795 & 7.15$\substack{ +0.68 \\ -0.79 }$ & 131.89$\substack{+6.32 \\ -7.33}$\\
A1991 & 111.22$\substack{ +0.83\\ -0.92 }$ & 11.15$\substack{+0.04 \\-0.04 }$\\ 
A2029  &  9.39$\substack{ +0.66\\ -0.76 }$ & 134.31$\substack{+4.72 \\-5.45 }$ \\
A2390 &  5.83$\substack{ +0.22\\ -0.23 }$ & 137.18$\substack{+2.60 \\ -2.81}$\\
RX J1159+5531 & 41.06$\substack{ +1.33\\-1.19 }$ & 34.07$\substack{+0.55 \\-0.49 }$\\
MKW 4 & 102.4$\substack{ +0.92\\-0.98 }$ & 10.31$\substack{+0.04 \\ -0.04}$\\
\end{tabular}
\end{ruledtabular}
\caption{\label{tab:table4} Estimated values for the core density ($\rho_c$) and the core radius ($r_c$)  for the V06 cluster sample.}
\end{table}

\begin{table}[h]
\begin{ruledtabular}
\begin{tabular}{ccc}
Slope & Intercept & Cluster Sample \\ \hline
$-1.47\pm 0.04$ & $0.75 \pm 0.08$ & ROSAT (single-$\beta$ profile) \\
$-1.46\pm 0.16$ & $0.88 \pm 0.33$ & ROSAT (double-$\beta$ profile) \\
$-1.30\pm 0.07$ & $0.6 \pm 0.11$ & ROSAT (cool-core clusters) \\
$-1.50\pm 0.24$ & $0.96 \pm 0.54$ & ROSAT (non cool-core clusters) \\
$-1.64\pm 0.1$ & $1.58 \pm 0.21$ & LOCUSS  \\
$-1.08^{+0.06}_{-0.05}$ & $0.4^{+0.24}_{-0.25}$ & Chandra (this work) \\

\end{tabular}
\end{ruledtabular}
\caption{\label{tab:summary}Summary of results for a linear regression of  $\ln(\rho_c)$ versus $\ln (r_c)$ from different cluster samples. All the other results are from V06.}
\end{table}

\subsection{Comparison with C14}
\label{sec:C14comparison}
Given the large discrepancies in the slope of our scaling relation with respect to C14, we investigate  the reason for these differences. As mentioned earlier, C14 assumed that the dark matter mass in cluster is the same as the total mass. This assumption is not correct, as the  gas mass fraction in the Chandra cluster sample ranges from (10-15)\%~\cite{2006ApJ...640..691V}.  Furthermore, 
C14 did not account for any intrinsic scatter in the scaling relation. Instead, the slope and intercept were determined using the BCES method~\cite{BCES}.
\rthis{Among the multiple implementations of the BCES technique~\cite{BCES}, C14 used the bisector method.}

To reconcile the differences between our result and C14, we now redo our analysis in exactly  the same way as C14, and also try a few variants to understand the impact of each of the different assumptions between our methods. Therefore, for this set of relaxed clusters, we also carried out the same analysis with the BCES\rthis{-bisector} method,   without subtracting the gas and stellar contribution as well as by subtracting these contributions. We implemented the BCES\rthis{-bisector} technique using the {\tt bces} module~\cite{BCES,2012Sci...338.1445N} in Python~\footnote{\url{https://github.com/rsnemmen/BCES}}.
We note that the  BCES method~\cite{BCES} does not account
for intrinsic scatter, when the abscissa contains errors~\cite{Ascenso}.  \rthis{Furthermore, the  authors of the {\tt bces} module have also cautioned  against  using the bisector method in BCES, as this method is self-inconsistent~\cite{Hogg}.}

Finally, we also did a fit with the same likelihood as in Eq.~\ref{eq:eq13},  without accounting for intrinsic scatter but by subtracting the gas and star mass,  and also vice-versa. \rthis{Again, we used the {\tt emcee}~\cite{emcee} module for carrying out the maximization of the log  likelihood.} A summary of the resulting slopes using these different combinations of assumptions can be found in Table~\ref{tab:table4}. 

When we implement the BCES method, and do not subtract the gas and star mass (which exactly duplicates the C14 procedure), our value for the slope $(-1.44 \pm 0.18)$ is consistent  with the value found by C14 for the ROSAT sample. Even when we maximize the log-likelihood, we get a value for the slope  $(-1.37 \pm 0.02)$  closer  
to that found in C14. With both the BCES method and the maximization of the log likelihood (without intrinsic scatter), we get a near constant value for the dark matter surface density, only when 
we subtract the gas and star mass. Finally, if we do not subtract the gas and star mass, but do a fit using Eq.~\ref{eq:eq11} to allow for intrinsic scatter, we get a slope \rthis{consistent with} -1 (within $1\sigma$).

Therefore, we conclude that we can reproduce nearly the same  value  for the slope of the regression relation between $\ln (\rho_c)$ and $\ln (r_c)$ as C14, when we use the BCES\rthis{-bisector} method (which does not fit for an intrinsic scatter), and 
also assume  that the total dark matter mass is the  same as the total cluster  mass. However,  to correctly estimate the scaling relations, one must subtract the gas and star mass estimates and also allow for an intrinsic scatter as we have done.

\begin{table}[h]
\begin{ruledtabular}
\begin{tabular}{cccc}
 Method & Gas/Star Mass  &  Slope   \\
 &Subtracted?&\\ \hline
BCES Method & No & $-1.44\pm0.18$  \\
BCES Method & Yes & $-1.09\pm0.02$  \\
W/o Intrinsic Scatter & No & $-1.37\pm0.02$  \\
W/o Intrinsic Scatter & Yes & $-0.99\pm 0.003$  \\
With Intrinsic Scatter & No & $-1.11\pm 0.29$ \\
\end{tabular}
\end{ruledtabular}
\caption{\label{tab:table3i}Summary of results for a linear regression of  $\ln\rho_c$ versus $\ln r_c$ using the Chandra X-ray sample by choosing combinations of different fitting methods (BCES\rthis{-bisector} vs maximizing Eq.~\ref{eq:eq11}) as well as other assumptions related to intrinsic scatter and subtraction of gas/star mass. When we use the BCES method and do not exclude gas/star mass, our result for the slope agrees with C14 for ROSAT cluster sample (cf. Table~\ref{tab:summary}).}
\end{table} 

\section{Comparison  with theoretical models}
\label{sec:theory}
Our results from the previous section show that we see deviations from a  constant halo surface density only at about 1.4$\sigma$. However, the resulting value of our  halo surface density is about ten times larger than for galaxies. We briefly discuss some theoretical models which are consistent with such a scenario.


The problems with $\Lambda$CDM at small scales can be solved by self-interacting dark matter with a velocity-dependent cross-section ranging  from $\sigma/m \approx 2 cm^2/ g$ on galaxy scales to 
$\sigma/m \approx 0.1 cm^2/ g$ on cluster scales~\cite{Kaplinghat15,Tulin}. C14 had argued  that such velocity-dependent self-interacting cross-sections for dark matter (SIDM) ~\cite{Vogelsberger,Weiner,Rocha,Kaplinghat15} are ruled out, if the observed core is produced due to dark matter self-interactions. The reason for this is
that their scaling relations using the ROSAT sample  showed that the halo surface density scales with $r_c$ as $r_c^{-0.46}$. However, combining the $r_c-V_{max}$ and
$\rho_c-V_{max}$ scaling relations in ~\cite{Rocha}, one gets that the  halo surface density scales with $r_c^{0.6}$, showing the opposite trend. However, as we showed in the previous section, C14's scaling relation was obtained using incorrect assumptions. Furthermore, one cannot rule out SIDM, by comparing only cluster-based halo scaling relations with simulations, since the dynamic range for $V_{max}$ using only clusters is not large enough to test the scaling relations over a wide range of values.

Using cosmological simulations of self-interacting dark matter, Rocha et al~\cite{Rocha} showed that
$\frac{\sigma}{m} \approx 0.1 cm^2/g$ correctly predicts cluster scale halo surface densities comparable to the observed ones in this work.
Kamada et al~\cite{Kaplinghat} pointed out that  for $\frac{\sigma}{m}$ of 3~$cm^2/g$, the halo surface density for galaxies scales as  $V_{max}^{0.7}$, and assuming $V_{max}$ in the range of 20-100 km/sec, one gets  values for the halo surface density for galaxies in the same ballpark as found by observations~\cite{Donato,Salucci19}. For galaxy clusters $V_{max}$ is usually about one to two orders of magnitude larger than for galaxies. Therefore, a simple extrapolation (along with a gradual decrease in cross-section as a function of velocity)   to cluster scales should provide the correct order of magnitude for the  halo surface density on cluster scales.  Of course, a more detailed comparison is beyond the scope of this work and one would need
to calculate the surface density using mock catalogs from the latest SIDM simulations.

Lin and Loeb~\cite{Loeb} proposed another semi-analytical model of self-interacting dark matter  involving dark matter annihilations, wherein they provided an empirical formula for the dark matter core halo surface  density as a function of halo mass (Eq.~\ref{loebeq}).  They showed that this halo density agrees  with galaxy scale observations~\cite{Donato, Salucci19} as well with the data for one galaxy cluster (Phoenix)~\cite{SPT2012}. Since our halo surface densities are approximately the same  as that for the Phoenix cluster, we argue that the model proposed in Ref.~\cite{Loeb} correctly predicts the  one order of magnitude increase  in the observed surface density, as we go from  galaxy-scale to cluster-scale haloes.  This model also predicts a mild dependence on $M_{200}$, which we test in the next section.

Del Popolo et al have argued in a series of works~\cite{DelPopolo12,DelPopolo14,DelPopolo17,DelPopolo20} that Eq.~\ref{delpopoloeq}, which can be explained within the context of the  secondary infall model~\cite{Delpopolo09} agrees with  halo surface density of galaxies as well  as clusters by comparing with the data for galaxies in Ref.~\cite{Donato,Gentile}
along with additional data sets~\cite{Napolitano,DelPopolo12,Saburova} (which they analyzed themselves), and also with galaxy  cluster data compiled in Ref.~\cite{Boyarsky}. They however fit the column density instead of the surface density, and find that 
 the column density shows a mild dependence on the galaxy mass and galaxy  magnitude, which can be easily explained using their model. We also note that Eq.~\ref{delpopoloeq} gives the right order of magnitude for the halo surface density, which we obtain for the Chandra cluster sample.

For other models within and beyond $\Lambda$CDM,  which explain a constant halo surface density on galactic scales~\cite{Ogiya,Famaey,Baushev,Matos,Berezhiani,Milgrom}, we could not find any definitive predictions therein for cluster scale halo surface densities.

Finally, we note that most recently, Burkert pointed out that for fuzzy dark matter models, the halo surface density scales inversely with the cube of the core radius and is therefore in complete disagreement with observations~\cite{Burkert20}. Hence, the constant dark matter surface density observations~\cite{Donato,Salucci19} are in conflict with fuzzy dark matter scenarios.

Therefore, we conclude that the results in this work in conjunction  with the   constant halo surface density values on galaxy scales, are consistent with velocity-dependent self-interacting dark matter models as well as the secondary infall model (within $\Lambda$CDM), but are in tension with fuzzy dark matter. For a more definitive test with various theoretical models one needs to extend this analysis to galaxy groups, which bridge the mass gap between galaxies and clusters, and then compare with predictions from simulations.


\section{Dependence on $M_{200}$}  
\label{sec:m200}
We now use our results for $\rho_c$ and $r_c$ to check for correlation with $M_{200}$, as suggested in some works~\cite{DelPopolo12,Loeb}.  The first step in doing this is to estimate $M_{200}$ from $M_{500}$. In V06, the masses ($M_{500}$) and concentration parameters ($c_{500}$) for the overdensity (with respect to the critical density)  level $\Delta=500$ and its corresponding radius ($r_{500}$) have already been derived, where $c_{500}$ was obtained using the mass-concentration relations in ~\cite{Dolag}. We have determined the $M_{200}$ values using same prescription as in Ref.~\cite{2017ApJ...844..101A}, which assumes an NFW profile
\begin{equation}
    M_{200}=M_{500}\frac{f(c_{200})}{f(c_{500})}
    \label{eq:eq15}
\end{equation}
where $f(c)$ is a function of the concentration $c$ and the over-density ($\Delta$), and is given by~\cite{2017ApJ...844..101A}
$$f(c_\Delta)=\ln(1+c_\Delta)-\frac{c_\Delta}{1+c_\Delta}$$ 
The concentration at $\Delta=200$,  $c_{200}$ was obtained by solving for the following equation
\begin{equation}
\frac{c_{200}^3}{\ln(1+c_{200})-\frac{c_{200}}{1+c_{200}}}=\frac{3\rho_s}{200\rho_{c,z}}
    \label{eq:eq16a}
\end{equation}
where the NFW density scale parameter ($\rho_s$) and scale radius were fixed for each cluster and is given by
$$\rho_s=\frac{M_{500}}{4\pi r_s^3} \quad \rm{with} \quad r_s=\frac{r_{500}}{c_{500}}$$
and $\rho_{c,z}$ is determined from
$$\rho_{c,z}=\frac{3M_{500}}{2000\pi r_{500}^3}$$
The $M_{500}$ values for the clusters A262 and MKW4 were unavailable in V06. Hence, we calculated it by extrapolating the mass profiles to the corresponding $r_{500}$ values. 
To calculate the error in $M_{200}$, we propagated the errors in $M_{500}$ and  $c_{500}$ provided in V06. 

\begin{figure*}
    \includegraphics{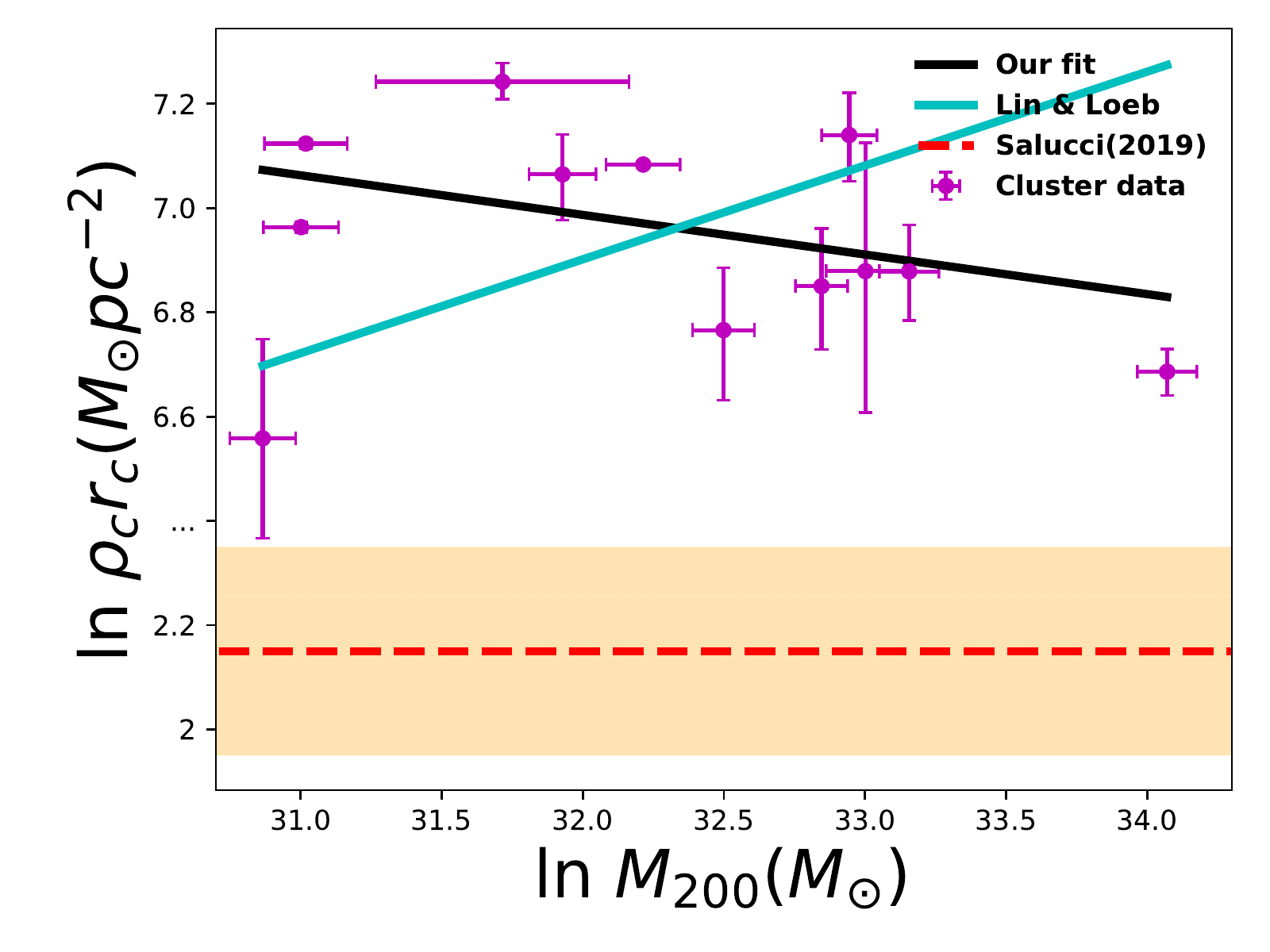}
    \caption{$\ln (\rho_c r_c)$ versus $\ln M_{200}$ from V06 cluster sample~\cite{2006ApJ...640..691V}. The units for $\rho_c r_c$ and $M_{200}$ are in $M_\odot pc^{-2}$ and $M_\odot$ respectively. The black line represents the fitted line of our analysis, whereas  the cyan line represents the models   from Lin \& Loeb~\cite{Loeb}. We get similar fit for the scaling relation predicted by Del Popolo et al~\cite{DelPopolo12}, which we have omitted from the plot for brevity. The red dashed line indicates the constant surface  obtained from single galaxy systems of various types~\cite{Salucci19},  while the orange shaded region represents 1$\sigma$ error. Note that the mass range for these systems is much lower than for clusters. Note that in this plot the range of  values 2.4-6.4 have been culled from the Y-axis in this plot, given the large difference in surface density for single galaxies and clusters.}
    \label{fig:f4}
    \end{figure*}

The relation between the dark matter column density, $S=\rho_c r_c$ and $M_{200}$ in logarithmic  space is shown in Fig \ref{fig:f4}.  
We have again done a linear regression with $y= \ln \rho_c r_c$ and $x=M_{200}$; and maximized the log-likelihood function (same as Eq.~\ref{eq:eq13}) using the {\tt emcee} MCMC sampler with uniform priors. The best-fit parameters thus obtained are,
\begin{eqnarray}
\ln\bigg(\frac{\rho_c r_c}{M_{\odot} pc^{-2}}\bigg)=(-0.07^{+0.05}_{-0.06}) \ln\bigg(\frac{M_{200}}{M_\odot}\bigg)\nonumber\\
+(9.41^{+2.07}_{-1.80})  
\label{eq:eq16} 
\end{eqnarray}

The intrinsic scatter for this fit is $17^{+4.0}_{-6.0} \%$.
We now fit this data to two scaling relations predicted by  two independent theoretical scenarios proposed in literature for the dark matter halo surface density. Del Popolo et al~\cite{DelPopolo14} found after applying the secondary infall model proposed in Ref.~\cite{Delpopolo09} to cluster data ~\cite{Boyarsky09}, that $S \propto M_{200}^{0.16}$, where $S$  is the dark matter column density.
Lin and Loeb deduced from numerical simulations of self-interacting dark matter with annihilations, that $S \propto M_{200}^{0.18}$, where $S$ is the product of the halo core density and radius~\cite{Loeb}. However, we should caution that although the definition of core density in ~\cite{Loeb} is same as ours, the core radius defined in Ref.~\cite{Loeb}  could differ from the definition in Burkert profile depending on the initial scale radius in the dark matter halo profile~\cite{Loeb}. Similarly $S$ in Eq~\ref{delpopoloeq} is the dark matter column density, whereas what we have estimated is the dark matter halo surface density. Although, both give about the same values for both cuspy and cored profiles~\cite{DelPopolo12}, they are not exactly the same quantities.
Nevertheless, we would like to test for   the correlation with $M_{200}$ as predicted in both these works. We note that in the Lin and Loeb model~\cite{Loeb},  there is also a slight dependence of the surface density as a function of redshift (See Fig. 2 of Ref.~\cite{Loeb}). 
However, since no analytic formula for the variation with redshift is provided, we do not account for this. 
We further point out  these two  relations are not exhaustive and other proposed scaling relations for the dark matter surface density as a function of halo mass are discussed in Ref.~\cite{DelPopolo12}.

However, when we compare our estimated surface density with $M_{200}$, we find a slight decrease in dark matter core density  with $M_{200}$, although the  decrease with $M_{200}$ is not significant (cf. Table~\ref{tab:summary2}).
Therefore, at face value our results would not be consistent with these predictions. To carry out a more definitive test, we now try to fit our data to these relations, by using the same slope (0.18 and 0.16) as predicted by these models, with only the intercept and intrinsic scatter as free parameters. We then do a model comparison with our best fits using AIC and BIC information criterion~\cite{Liddle07}. AIC and BIC are defined as follows~\cite{Liddle07}:
\begin{eqnarray}
\rm{AIC} &=& -2\ln L_{max} + 2p + \frac{2p(p+1)}{N-p-1} \\
\rm{BIC} &=& -2\ln L_{max} + p \ln N, 
\end{eqnarray}
\noindent where $N$ is the total number of data points, $p$ is the number of free parameters in each model, and $L_{max}$ is the maximum likelihood. When comparing two models, the model with the smaller value of AIC and BIC is considered the favored one.

The best-fit results for these two scaling relations can be found in Table~\ref{tab:summary2}.
We find that $\Delta$AIC and $\Delta$BIC between our best-fit and that for other scaling relations is between 6-7, wherein our fit has the lowest value, indicating 
strong evidence for our fit as compared to the relations proposed in Refs.~\cite{DelPopolo12,Loeb}. We note that we shall obtain poorer fits for other scaling relations, which predict a steeper dependence with halo mass, for eg.~\cite{Boyarsky09}. We should however caution that the dynamical range in mass for our cluster sample
is not large, and for a more definitive test, lower mass samples should be included.

A comparison of our best-fit along with a fit to the theoretical relation in Lin and Loeb~\cite{Loeb} can be found in Fig.~\ref{fig:f4}. In the same figure, we also show for comparison the constant value for the surface density, obtained for single galaxy systems  using the latest data~\cite{Salucci19}.

\begin{table}[h]
\begin{ruledtabular}
\begin{tabular}{cccccc}
 Model &  Slope & $\sigma_{int}$ & AIC & BIC \\ \hline
Lin \& Loeb~\cite{Loeb} & 0.18 & 28\% &  12.7 & 14.7 \\
Del Popolo et al~\cite{DelPopolo12}&  0.16 & 26\% &  11.6 & 13.6  \\
This work & $-0.07^{+0.05}_{-0.06}$ & 17\% &  5.1 & 8.0  \\

\end{tabular}
\end{ruledtabular}
\caption{\label{tab:summary2}Summary of results for a linear regression of  $\ln(\rho_c r_c)$ versus $\ln (M_{200})$ from different models and their comparison using AIC and BIC. Our best-fit (Eq.~\ref{eq:eq16}) has the smallest values of AIC and BIC and the difference between the other two scalings is between 6-7 indicating strong preference for our model compared to the other two.}
\end{table}

\section{Conclusions}
\label{sec:conclusions}
A large number of studies in the past decade have found that the dark matter surface density, given by the product of dark matter core radius ($r_c$) and core density ($\rho_c$) is constant for a wide range of galaxy systems from dwarf galaxies to giant galaxies  over 18 orders in blue magnitude.  This cannot be  trivially predicted by the vanilla $\Lambda$CDM model, but it can be easily accommodated in  various alternatives to $\Lambda$CDM or by invoking various feedback mechanisms in $\Lambda$CDM.

However, there have been very few tests of this {\it ansatz} for galaxy clusters. The first systematic study of this relation for a large X-ray selected cluster sample was done by C14 using the ROSAT sample studied in Chen et al~\cite{Chen07}. They considered a sample of galaxy clusters in hydrostatic equilibrium  and using parametric models for gas density and temperature, obtained the total mass density profile. They assumed that this is a proxy for the total dark matter density distribution. For this sample, $\rho_c$ was obtained by extrapolating the dark matter density distribution to the center of the cluster, whereas, $r_c$ was obtained by determining the radius at which the core density drops by a factor of four. This emulates the definition of core radius in the Burkert cored dark matter profile~\cite{Burkert95}. Therefore, this analysis was done without positing any specific dark matter density distribution.
C14 did not find a constant dark matter surface density, but found a tight scaling relation between $\rho_c$ and $r_c$, given by $\rho_c \propto r_c^{-1.46 \pm 0.16}$.

We then carried out a similar analysis as in C14 for a Chandra X-ray sample of 12 relaxed clusters, for which detailed 3-D gas density and temperature profiles were made available by Vikhlinin et al~\cite{2006ApJ...640..691V}. One improvement on the analysis in C14, is that we also subtracted the gas and star mass, while non-parameterically reconstructing the dark matter density profile. Furthermore, while determining the scaling relations between the core density and radius, we also accounted for the intrinsic scatter. 
Our results for the dark matter core density and radius can be found in Table~\ref{tab:table4}. They are of the same order of magnitude as previous estimates for galaxy clusters~\cite{Chan}, and are about an order of magnitude larger than for isolated galaxy systems. The halo surface densities for the cluster scale haloes  are in the right ballpark from secondary infall model~\cite{Delpopolo09}, as well as velocity-dependent self-interacting dark matter scenarios~\cite{Tulin}.

We  find that $\rho_c \propto r_c^{-1.08^{+0.06}_{-0.05}}$. The intrinsic scatter for this relation is about 18 \%. Therefore, we get only a marginal deviation from a reciprocal relation between the dark matter core
density and radius in contrast to C14 who found a steeper dependence of $\rho_c$ as a function of $r_c$. Our estimated dark matter surface density  is inconsistent with flat density core at only  $1.4\sigma$.  A comparison of our result with previous scaling relations found for galaxy clusters can be found in Table~\ref{tab:summary}. We also checked that the discrepancy in our results compared  to C14, is because C14 did not subtract the gas and star mass, or account for an intrinsic scatter. If we replicate exactly the same procedure as C14, we can reproduce their scaling relations using the Chandra cluster sample.

We also checked for any dependence of the dark matter surface density with $M_{200}$ to test some of these predictions in literature~\cite{DelPopolo12,Loeb}, although the exact definition in these works is not the same as the product of the dark matter core radius and density, which we calculated.
We find that
the dark matter surface density ($S$) scales with $M_{200}$ as $S  \propto M_{200}^{-0.07 \pm 0.55}$, which is in mild disagreement the weak logarithmic increase with  $M_{200}$ predicted in Refs.\cite{DelPopolo12,Loeb}.
However, a more definitive test can only be confirmed using a larger sample covering a wide dynamic range in mass by extending this test to galaxy groups.

Further stringent tests of this relation for clusters should soon be possible, thanks to the recent launch of the  e-ROSITA satellite, and the expected discovery of about 100,000 clusters~\cite{Hofmann}.
\section*{Acknowledgements}
We are grateful to Man-Ho Chan and Antonio Del Popolo for useful correspondence,  and Alexey Vikhlinin for providing us the data in V06. We are also thankful to the anonymous referee for several constructive feedback on our manuscript.
\bibliography{new}
\end{document}